\documentclass[useAMS,usenatbib]{mn2e}

\newcommand{\nodata} {$^{...}$}
\input{epsf}

\title[Rest-frame optical continua of $L \sim L^{*}$ $z>3$ quasars]
{Rest-frame optical continua of $L \sim L^{*}$, $z>3$ quasars: probing
the faint end of the high $z$ quasar luminosity function}
\author[O. P. Kuhn]
       {Olga P. Kuhn\\
        Joint Astronomy Centre, 660 N. A`ohoku Place, Hilo, HI, 96720 \\
{\tt email: o.kuhn@jach.hawaii.edu}
	}

\begin{document}

\maketitle

\label{firstpage}

\begin{abstract}
Near-IR photometry for 20 radio-loud $z>3$ quasars,
16 of which are radio-selected, are presented.
These data sample the rest-frame optical/UV continuum, which is commonly
interpreted as emission from an accretion disk.
In a previous study, we compared the rest-frame optical/UV
continuum shapes of 15 optically bright (V$<17.5$) $z>3$ quasars with
those of 27 low redshift ($z\sim0.1$) ones that were matched to the high 
redshift sample in {\it evolved} luminosity (i.e. having luminosities 
ranging from 1-7 times the characteristic luminosity, $L^{*}$, where
$L^{*} \sim (1+z)^{\sim3}$) to look for signs of evolution in the central
engines.
We found the continuum shapes at $z\sim0.1$ and $z>3$ similar, consistent with
no significant change in the ratio $\dot{m}$/M, where $\dot{m}$ is the accretion rate with
respect to the Eddington rate and M is the black hole mass.
This study expands our earlier high redshift sample to lower luminosity, away
from extreme objects and towards a luminosity overlap with lower redshift samples.
The distribution of rest-frame optical/UV continuum shapes for
this fainter sample is broader, extending further to the red than that of 
the brighter $z>3$ one. Three quasars from this fainter sample, two 
radio-selected and one optically-selected, have optical continuum slopes $\alpha < -1$ 
($F_{\nu} \sim \nu^{\alpha}$). 
The optically-selected one, LBQS0056+0125, appears to be reddened
by dust along the line of sight or in the host galaxy, whereas the radio-selected
ones, PKS2215+02 and TXS2358+189, could derive their red continua from the contribution
of a relatively strong synchrotron component to the rest-frame optical.
These objects may represent a bridge to a population of very red
high redshift quasars to which ongoing or future near-IR, optical and 
deep X-ray surveys will be sensitive.
\end{abstract}


\section[]{Introduction}

Within the past decade, there has been significant progress towards
characterizing the evolution of the quasar luminosity function.
Up to redshifts $z<2$, it is well described
as luminosity evolution, with the characteristic luminosity $L^{*}$
increasing by a factor $40-50$ from $z=0$ to $z=2.3$ (Boyle et al. 2000).
Quasar activity peaks around z$\sim2.5-3$ (e.g. Schmidt, Schneider \& Gunn 1991)
but from there to $z>4$, the space density drops by a factor $\sim2-20$
(e.g. Jarvis \& Rawlings 2000, Fan et al. 2001a, Vigotti et al. 2003).
This high redshift turnover is identified with the epoch of quasar formation
(Warren, Hewett \& Osmer 1994), but the
cumulative effects of dust along the line of sight may also play a role, the significance
of which is not yet well determined.

Despite our improved knowledge of the statistical evolution of quasars,
a fundamental question, `how do individual quasars form and evolve?', remains  
unanswered. Models which combine theories of structure formation with those of
energy generation in quasars can reproduce the evolution of the quasar
luminosity function (e.g. Siemiginowska \& Elvis 1997; Haehnelt, Natarajan \& Rees 1998; 
Haiman \& Menou 2000; Kauffman \& Haehnelt 2000; Hatziminaoglou, Siemiginowska \& Elvis 2001), 
but these differ in details so do not constrain the evolution of the central engine.

The optical/UV continua and emission lines originate within the central
parsec and reflect conditions in the central engine. At low redshift,
the emission line correlations involved in Boroson \& Green's (1992) 
eigenvectors 1 and 2 have been well studied, and within the past couple of
years in particular, the physical drivers of these have become better
understood, enabling them to be used as tracers of the central mass and accretion rate 
(Marziani et al. 2001; Boroson 2002).
The optical/UV continuum is often attributed to emission from an
accretion disk. Model fits can be used to estimate the mass and accretion
rate, though factors such as intrinsic reddening or additional components,
such as synchrotron emission which may be important in the
flat-spectrum radio quasars (Serjeant \& Rawlings 1997; 
Francis, Whiting \& Webster 2000; Whiting, Webster \& Francis 2001),
complicate matters.
Use of multiple datasets: emission line, continuum and polarization; 
should yield the best estimates of central engine parameters, however, these
are feasible only for the brighter quasars.

In a previous study, we compared the rest-frame optical/UV continua
of 15 $z>3$ quasars with those of 27 $z\sim0.1$ quasars matched
to the high redshift ones in {\it evolved} luminosity (Kuhn et al. 2001, hereafter
Paper I). We found no evidence for significant evolution of the continuum shape.

The high redshift sample in our previous study was selected nearly a 
decade ago and was limited. We had aimed to gather spectroscopic as
well as photometric data. This meant that the objects had to be bright
and relatively few. The resulting set which consisted of the optically brightest $z>3$ 
quasars was nearly evenly split between radio-loud and radio-quiet quasars, and several of 
these were considered the most luminous objects in the universe when they were discovered (e.g.
HS 1946+7658; Hagen et al. 1992). Finally, our low-z/high-z sample matching criteria imposed
a strong redshift-luminosity degeneracy.

To move the high-redshift sample away from the most extremely luminous 
objects, towards some overlap with the low-redshift one (and other
low or intermediate redshift samples drawn from the literature), and
increase its size, a set of 20 fainter $z>3$ quasars was selected. 
For the previous `bright $z>3$' sample, a comparison of 
the optical/UV spectral indices measured using both photometric and 
spectroscopic data with those measured from the photometry alone 
shows that photometry alone is adequate at least for
single power law fits to continuum shape.
This paper describes the observations and distribution of optical/UV continua
for this new `faint $z>3$' sample. Section 2 discusses the sample
selection and section 3, the data acquisition and reduction. In section
4, the rest-frame optical/UV continua are presented and, in section 5, 
compared with those of the `bright $z>3$' sample.
Section 6 discusses the reddest quasars with $\alpha<-1$, and conclusions are drawn 
in section 7. 
Except in Figure \ref{zldist}, a Friedmann cosmology with 
H$_{o}=75$ km s$^{-1}$ Mpc$^{-1}$, q$_{o}=0.1$ and $\Lambda=0$ is used.

\section[]{Sample}

\subsection{Selection}
To minimize optical selection effects, the
`faint $z>3$' sample was limited to radio-loud quasars. 
The targets were drawn from a listing of all objects with $z>3$ 
that were catalogued in NED\footnote{NED, the NASA/IPAC Extragalactic 
Database} as both quasars and radio sources. To be 
accessible from UKIRT and from La Palma (where we sought optical data), 
a declination limit: $-20^{\circ} < \delta < 60^{\circ}$; was imposed. 
A cross-check with the latest edition of the V\'eron-Cetty \& V\'eron 
catalogue (2000; VCV9) revealed a few new quasars; and vice-versa, a few NED 
quasars were not in VCV9. Quasars appearing in one or the other catalogue 
were kept. Those noted as lensed were removed, and a final list of 20 
radio-loud quasars accessible during the allocated nights, i.e. with right
ascension between 22$^{\rmn{h}}$ and 08$^{\rmn{h}}$30$^{\rmn{m}}$, 
constituted the target sample, hereafter referred to as {\em z3f} (for $z>3$
faint; Table \ref{z3fsample}).

Of the 20 {\em z3f} quasars, 16 were radio-selected and 4 were originally discovered
in optical | objective prism or grism | surveys. The radio-selected quasars tend to have 
high radio-to-optical flux ratios (Table \ref{z3fsample}) and flat radio spectra 
(Table \ref{alphas}). This is probably a consequence of the survey frequency (most are from 5GHz
and 1415MHz surveys, so sample still higher rest-frame frequencies) and sensitivity. 
Deeper and longer wavelength surveys would be more effective in selecting steep-spectrum
high redshift quasars.
This radio-selected subsample avoids the bias towards blue optical/UV continua that may result from
some optical surveys, but on the other hand may include blazars in which a synchrotron component
contributes to and reddens the optical/UV continua (Whiting et al. 2001). The 4 optically selected 
radio-loud quasars were kept so that their colors could be compared with those of the 
radio-selected ones.

\subsection{Radio-loudnesses}
The sample quasars are classified as radio-loud no matter which criterion |
rest-frame radio-luminosity or radio-to-optical flux ratio | is used.
Their rest-frame 1.4GHz luminosities are computed using equation 1, from Stern et al.
(2000), where I have assumed the radio spectral index, 
$\alpha_{r}=-0.5$ ($F_{\nu}\sim\nu^{\alpha_{r}}$), to be consistent with the assumptions 
made by Gregg et al. (1996) who classify as radio-loud those quasars with 
log($L_{1.4GHz}$)$>32.5$ 
[erg s$^{-1}$ Hz$^{-1}$] (for H$_{o}=50$, q$_{o}=0.5$). 
%
\begin{equation}
L_{1.4GHz} = {4 \pi d_{L}^{2}} {{F_{1.4GHz,obs}}\over {(1+z)^{(1+\alpha_{r})}}}
\end{equation}
In the equation, $d_{L}$ is the luminosity distance and $F_{1.4GHz,obs} (\equiv F_{NVSS}$) is the 
observed flux at 1.4 GHz from the NRAO-VLA Sky Survey (NVSS; Condon et al. 1998). 
The values of $F_{1.4GHz,obs}$ and $L_{1.4GHz}$ are listed in the fourth and fifth columns
of Table \ref{z3fsample}. To compare the luminosities $L_{1.4GHz}$ with the radio-quiet/radio-loud 
threshold of Gregg et al. (i.e. to convert them to the cosmological parameters that Gregg et al. adopted), 
a constant which depends on redshift but ranges from $0.02$ to $0.1$ for this sample should be 
subtracted. All of the {\it z3f} quasars have rest-frame 1.4 GHz luminosities 
above the radio-loud threshold of Gregg et al (1996).
The radio-to-optical flux ratio, RL, is calculated as in Bechtold et al. (1994), 
although the rest-frame 5GHz and 1450\AA\ luminosities are computed directly from the 
radio data available from NED and the NVSS and from the near-IR data published here: 
\begin{equation}
RL = \log_{\tiny 10} {L_{5GHz} \over L_{1450\AA}} 
\end{equation}
Objects with $RL>1$ are considered radio-loud.
For the three objects with only one radio measurement, a radio spectral index, $\alpha_{r} =
-0.18$ is assumed. This is the median $\alpha_{r}$ for the 16 radio-selected quasars. 
Values of RL for the sample quasars are listed in column 6 of Table \ref{z3fsample}.

\subsection{Radio spectral indices}
\label{rspec}

Most of the quasars were classified as flat-spectrum radio-quasars. The spectra of a
couple which had more extensive radio data were determined to be peaked at Giga-Hz
or higher frequencies (as noted in Table \ref{z3fsample}). One, B30749+420, 
was classified as a compact symmetric object (CSO).
To measure their radio spectral indices, I fit a power law to the radio data taken from NED.
The results confirm that none of the quasars with greater than one
radio-flux measurement, i.e. the 16 radio-selected {\em z3f} quasars plus Q$2311-036$, is a steep-spectrum quasar; 
all have $\alpha_{r} > -0.5$ (column 6 of Table \ref{alphas}). 
Radio data at a greater range of wavelengths would be needed to determine whether any 
more of these are Giga-Hz peaked or CSOs. The tendency to find GPS quasars at high redshift
and their interpretation as young radio sources (O'Dea 1998) suggests that there may be more 
within this high redshift sample.

\subsection{$z-$L coverage}
\begin{figure}
\includegraphics{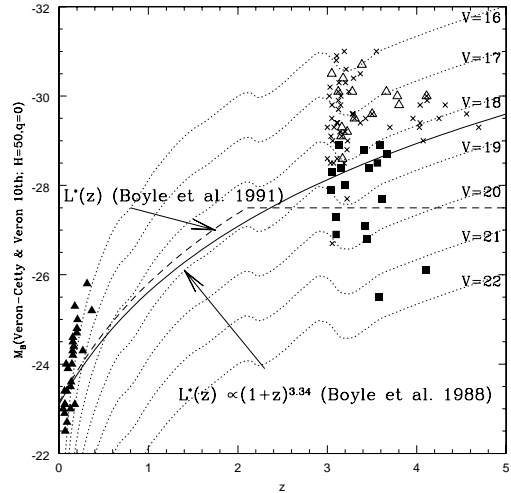} 
\vspace*{6.5cm}
\caption{Redshift-absolute magnitude distribution of three $z>3$ datasets considered here:
(1) 17 of the 20 target ({\it z3f}) quasars that are listed in, so have absolute magnitudes
from, V\'eron-Cetty \& V\'eron (2001, VCV10; filled boxes);
(2) the 15 previously studied bright $z>3$ quasars (open triangles); and (3) 
56 quasars with JHK$_{s}$ from 2MASS (crosses).
For reference, the redshift-absolute magnitude distribution of the 27 `Atlas' (Elvis et al. 1994a) 
quasars used as a comparison sample in Paper I are also plotted (filled triangles).
All the absolute magnitudes are from VCV10, which assume a k-correction based on a 
$F_{\nu}\sim\nu^{-0.3}$ approximation to the quasar continuum. 
The solid curve traces characteristic luminosity as a function of redshift:
$L^{*} \sim (1+z)^{3.34}$ up to $z\sim2$ (Boyle, Shanks \& Peterson 1988); the dashed 
line shows a later formulation which takes into account the slowing of evolution at $z\sim2$
(note that in both cases, the curve is only an extrapolation for $z>\sim2$).
Dotted lines mark tracks of constant apparent magnitude.
A Friedmann cosmology with H$_{o}$=50 km s$^{-1}$ Mpc$^{-1}$ and q$_{o}=0$ is
used for this plot, since that is what VCV10 adopt when computing absolute magnitudes and
it is one of the cosmologies for which Boyle et al. (1988, 1991) parameterize luminosity evolution.
\label{zldist}}
\end{figure}

A consideration in selecting the sample was to increase the luminosity
range at high redshift and reduce the redshift-luminosity degeneracy that 
was inherent in our earlier study. Figure \ref{zldist} plots the redshift-absolute magnitude
distribution of the `faint $z>3$' sample, together with the earlier
`bright $z>3$' (hereafter {\em z3b}) and low redshift ones.

Cross-correlating the $z>3$ quasars in the VCV9 
catalog with the positions of 2MASS sources in the 2nd incremental data 
release\footnote{{\em{http://www.ipac.caltech.edu/2mass/releases/second}}}  
yields 56 with detections at J, H and Ks. 
The redshifts and absolute magnitudes of these 2MASS-detected quasars are also indicated 
Figure \ref{zldist}.  While not homogeneous, the sample is large 
and as such provides a reference for the other $z>3$ samples.

\begin{table*}
\centering
\begin{minipage}{140mm}
\caption{Faint $z>3$ radio-loud sample} \label{z3fsample}
\begin{tabular}{lllrllll}
Quasar$^{a}$ &  z$^{a}$ & $m^{b}$ & F$_{NVSS}^{c}$ & L$_{1.4GHz}^{d}$ & RL$^{e}$ & radio surveys$^{f}$ & notes$^{g}$\\
PKS 2215+02      &  3.55 &21.5$^{h}$&781& 35.6 & 4.6& 87GB, TXS, WB92, PHJFS & FSRQ\\
MG3 J222537+2040 &  3.56 &  R18.6  & 221& 35.1 & 3.0& 87GB, WB92 &  FSRQ\\
MG3 J225155+2217 &  3.668&   20.2  & 190& 35.1& 3.9& 87GB, WB92& FSRQ \\
Q2311-036        &  3.034& R18.8 & 79 & 34.5 & 2.9& PMN, opt(UM 659) & FSRQ,ALS(1)\\
TXS 2342+342     &  3.053&   18.4  & 155& 34.8 & 2.9& 87GB, WB92 & FSRQ,ALS(2) \\
TXS 2358+189     &  3.10 &   20.5  & 266& 35.0 & 4.5& 87GB, WB92 & FSRQ,ALS(2) \\
MG1 J000655+1416 &  3.20 &   $-$   & 184& 34.9 & 2.9& 87GB, WB92, TXS& FSRQ\\
PC0027+0525      &  4.099&  R21.49 &   5& 33.6 & 3.1 & opt(SSG) & \\
LBQS0056+0125    &  3.149&   18.6  &   6& 33.4 & 1.9&  opt(SSG,LBQS) & ALS(3)\\
UM672            &  3.119& 18    &  46& 34.3 & 2.5&   opt(UM) & \\
MG3 J015105+2516 &  3.10 &  R19.8  & 200& 34.9 & 3.9& 87GB, WB92 & FSRQ\\
PKS 0201+113     &  3.61 &  R19.5  & 781& 35.7 & 3.9& 87GB, WB92, TXS & GPS(5),FSRQ(6),ALS(2)\\
MG3 J023222+2318 &  3.42 &  R19.9  & 461& 35.4 & 4.1& 87GB, WB92, TXS & FSRQ\\ 
MG1J024614+1823  &  3.59 &   $-$   & 217& 35.1 & 3.5& 87GB, WB92 & FSRQ\\
PKS 0335-122     &  3.442& R20.2 & 476& 35.4 & 4.5& PMN, TXS & FSRQ, ALS(4)\\
PKS 0336-017     &  3.197& R18.8 & 593& 35.4 & 3.7& PMN, 87GB, WB92, TXS, PHJFS & GPS(7)\\
MG2 J062425+3855 &  3.469&   R18.7 & 809& 35.6 & 4.2& B3, 87GB, WB92, S4, TXS & FSRQ\\
Q0642+449        &  3.396&  18.49  & 453& 35.4 & 3.8& B3, 87GB, WB92, S4, TXS & HFP(8)\\
B3 0749+426      &  3.59 &   R18.3 & 711& 35.6 & 3.2& 87GB, WB92, TXS  & CSO(9)\\
PMNJ0833+0959    &  3.75 &   $-$   & 122& 34.9 & 3.1& 87GB, WB92 & FSRQ\\
\end{tabular}
\end{minipage}
\begin{minipage}{140mm}
{\it a~}{Name as it appears in NED: coordinates used were from NED and not repeated here. 
Redshift $z$ from NED.} \\
{\it b~}{optical magnitude, at V if not indicated as R, from VCV10, 
except when otherwise noted.} \\
{\it c~}{F$_{NVSS}$ is the 21-cm flux in mJy from the NRAO-VLA Sky Survey (NVSS; Condon et al. 1998).}  \\
{\it d~}{Column lists $\log_{10}$(L$_{1.4GHz}$): L$_{1.4GHz}$ is the rest-frame luminosity at 1.4GHz in erg s$^{-1}$ Hz$^{-1}$.}  \\
{\it e~}{RL is the rest-frame radio(5GHz) to optical(1450\AA) flux ratio, as defined in
Bechtold et al. (1994); RL$>1$ is radio-loud. The radio and optical rest-frame fluxes are 
computed directly for each object from power law fits to the radio and to the near-IR data.} \\
{\it f~}{Radio and/or optical surveys which detected the quasar (indicated in this column and by the prefix in column 1):
PKS: 2.7 GHz, Parkes radio survey (e.g. Shimmins et al. 1966); 
PHJFS: PKS Half-Jansky Flat-Spectrum Sample (Drinkwater et al. 1997); 
MG: 5 GHz MIT Green Bank 5GHz survey (e.g. Bennett et al. 1986);  
87GB: 5 GHz (Gregory \& Condon 1991; Becker, White \& Edwards 1991);
TXS: 365 MHz (Douglas et al. 1996);
WB92: 1.4 GHz (White \& Becker 1992);
B3: 408 MHz, 3rd Bologna Catalog of Radio Sources (Ficarra, Grueff \& Tomassetti 1985);
S4: 5 GHz, Fourth `Strong' radio source survey (Pauliny-Toth et al. 1978);
PMN: 5 GHz, Parkes-MIT-NRAO radio survey (Griffith et al. 1994).} \\
{\it opt~} =discovered in an optical survey: 
LBQS -- objective-prism-selected (LBQS0056+0125 reported by Chaffee et al. 1991);
UM -- Univ. of Michigan, obj-prism selection (McAlpine \& Feldman 1982);
SSG -- grism survey (Schmidt, Schneider \& Gunn 1987, Schneider, Schmidt \& Gunn 1991). \\
{\it g~} Notes: \\
{\it ALS~} = line-of-sight absorption line system:
(1) Ly$\alpha$+metal system at $z\sim2.7$ (Bechtold 1994);  
(2) Ly$\alpha$ absorption systems at high redshift ($z>\sim3$) reported by White, Kinney \& 
Becker (1993). The one toward TXS 2358+189 may not be damped, but those toward TXS 2342+342 and 
PKS 0201+113 are; 
(3) $z\sim2.7$ damped Ly$\alpha$ system (Pettini et al. 1997);
(4) $z\sim3.178$ damped Ly$\alpha$ system (Ellison et al. 2001). \\
FSRQ = flat spectrum radio quasar; \\
GPS = Giga-Hz Peaked: (5) O'Dea, Baum \& Stanghellini (1991); 
(6) Stanghellini et al. (1998); 
(7) Savage et al (1990); 
(8) HFP = High Frequency Peaker (Dallacasa et al. 2000); 
(9) CSO = Compact Symmetric Object (Peck \& Taylor 2000). \\
{\it h~}{Optical magnitude from NED} \\
\end{minipage}
\end{table*}

\section[]{Observations \& Data Reduction}
\subsection{Near-IR photometry}

Near-infrared photometry was obtained for 18 of the {\it z3f} targets
in the fall of 2000 using UFTI (UKIRT Fast-Track Imager,
Roche et al. 2002) at the United Kingdom Infrared Telescope (UKIRT).
Data for the remaining 2 were obtained in the spring of 2001 using IRCAM at UKIRT.
Deeper images of the field of PC0027+0525 were obtained in August 2003,
and magnitudes measured from these rather than from the November 2000 ones were 
used in the anaylsis.
Table \ref{z3fir} lists the dates of the observations and results.
All 20 quasars were observed through the J, H and K {\it Mauna Kea Observatory NIR 
filters} ({\it MKO-NIR}; Tokunaga, Simons \& Vacca 2002), and 8 were also observed through the Z 
($\lambda_{cen} \sim 0.95\mu$m) filter.
Most of the observations were made on 17 and 18 November (UT), though a few
objects were observed in September and October to be quasi-simultaneous
with optical data which had been obtained in late August (Appendix A). All 20 were
re-observed approximately one year later (October 2001 - July 2002) 
at K to provide a rough indication of their variability (see 
section \ref{var}).

\begin{table*}
\scriptsize
\centering
\begin{minipage}{140mm}
\caption{Near-IR observations}\label{z3fir}
\begin{tabular}{lllllll} 
Quasar & Z$^{a}$ & J$^{a}$ & H$^{a}$ & K$^{a}$ & UT date & comments \\
PKS2215+02   & $19.63\pm0.05$ & $18.64\pm0.04$ & $18.05\pm0.03$ & $17.06\pm0.02$ & 2000 Sep 17 & Z-band data non-photometric \\ 
             &     \nodata      &     \nodata      &     \nodata      &$17.10\pm0.02$  & 2001 Oct 6 & $\Delta K = +0.04$ \\ 
%
MG3J222537+2040 & $17.69\pm0.04$ & $17.08\pm0.03$ & $16.70\pm0.04$ & $16.13\pm0.02$ & 2000 Sep 17 &  \\ 
                &  \nodata         &  \nodata         &  \nodata   & $16.07\pm0.03$ & 2001 Oct 6 & $\Delta K = -0.06$ \\ 
%
%
MG3J225155+2217     & \nodata &  $19.34\pm0.07$ & $18.67\pm0.06$ & $18.16\pm0.05$ & 2000 Nov 18 &  \\ 
                    & \nodata & \nodata &\nodata  & $17.98\pm0.04$ & 2001 Oct 6 & $\Delta K= -0.18$\\ 
Q2311-036& \nodata        &  $17.48\pm0.04$ & $16.76\pm0.03$ & $16.18\pm0.04$ &2000 Nov 17&  \\
         & $18.02\pm0.03$ &   \nodata   &   \nodata  &    \nodata  & 2000 Nov 18 &  \\ 
          & \nodata       &   \nodata   &   \nodata  &   $16.23\pm0.11^{b}$ & 2002 Jul 17 & $\Delta K= +0.05$ \\
%
TXS2342+342& \nodata & $17.26\pm0.02$ & $16.76\pm0.02$ & $16.34\pm0.01$ &2000 Oct 13 &  \\ 
           & \nodata & \nodata  & \nodata  & $16.34\pm0.02$  &2001 Oct 6 & $\Delta K = 0$  \\ 
TXS2358+189 & \nodata & $19.70\pm0.06$ & $18.76\pm0.04$ & $17.89\pm0.04$ & 2000 Nov 18 & scaled to Nov 17 \\
            & \nodata & $19.69\pm0.06$ & \nodata & \nodata & 2000 Oct 13 & \\ 
            & \nodata &  \nodata  & \nodata & $18.44\pm0.06$ & 2001 Oct 6 & $\Delta K= +0.55$ \\ 
MG1J000655+1416 & \nodata & $17.16\pm0.03$ & $16.98\pm0.04$ & $16.49\pm0.04$ &2000 Nov 17 &  \\ 
                & \nodata & \nodata & \nodata & $16.48\pm0.03$  &2001 Oct 8 & $\Delta K \sim 0$ \\ 
%
PC0027+0525 &\nodata & $20.55\pm0.09$ & $19.81\pm0.08$ & $19.10\pm0.10$ & 2000 Nov 18 & scaled to Nov 17 \\ 
            &\nodata & $20.46\pm0.04$ & \nodata        & $19.05\pm0.04$ & 2003 Aug 08 & $^{c}$ \\ 
%
LBQS0056+0125   &  \nodata  & $17.24\pm0.05$ & $16.54\pm0.05$ & $15.66\pm0.04$ & 2000 Nov 17&  \\ 
                &  \nodata  & \nodata & \nodata & $15.66\pm0.06$ & 2001 Oct 8& $\Delta K = 0$ \\ 
%
UM672           &  \nodata  & $16.99\pm0.04$ & $16.41\pm0.04$ & $15.75\pm0.04$ & 2000 Nov 17& \\ 
                &  \nodata  & \nodata        & \nodata        & $15.74\pm0.11^{b}$ & 2002 Jul 17& $\Delta K\sim0$  \\
MG3J015105+2516 &  \nodata  & $19.03\pm0.04$ & $18.72\pm0.05$ & $17.78\pm0.04$ & 2000 Nov 17& \\ 
                &  \nodata & \nodata & \nodata & $17.78\pm0.04$ & 2001 Oct 6& $\Delta K=0$ \\ 
PKS0201+113 &\nodata &$17.95\pm0.03$ & $17.54\pm0.03$ & $16.80\pm0.03$ & 2000 Nov 17&  \\ 
            &\nodata & \nodata  &  \nodata   & $15.65\pm0.03$ & 2001 Oct 6& $\Delta K = -1.15$ \\ 
%
MG3J023222+2318 &  \nodata  & $18.47\pm0.03$ & $17.94\pm0.04$ & $17.26\pm0.03$ & 2000 Sep 17&   \\ 
                &  \nodata  & \nodata &    \nodata     & $17.40\pm0.03$ & 2002 Feb 16& $\Delta K = +0.14$  \\ 
MG1J024614+1823 &  \nodata  & $17.86\pm0.03$ & $17.33\pm0.03$ & $16.30\pm0.05$ & 2000 Nov 17& \\ 
                &  \nodata  &   \nodata        &   \nodata  & $16.24\pm0.06$ & 2001 Oct 6& $\Delta K = -0.06$ \\ 
%
PKS0335-122 & $19.50\pm0.03$ & $19.04\pm0.05$ & $18.50\pm0.03$ & $17.56\pm0.03$ & 2000 Nov 18 & JHK scaled to Sep 17 \& Nov 17 \\ 
            &   \nodata & \nodata & \nodata & $17.75\pm0.04$ & 2002 Feb 16 & $\Delta K = +0.19$\\
%
PKS0336-017  &  $18.01\pm0.04$ &$17.54\pm0.04$ &$17.21\pm0.03$ &$16.65\pm0.03$ &2000 Nov 18 & \\ 
            & \nodata & $17.56\pm0.04$ &\nodata & \nodata & 2000 Oct 12 & \\ 
            & \nodata & \nodata &\nodata & $16.52\pm0.05$ & 2001 Oct 5 & $\Delta K = -0.13$ \\ 
MG2J062425+3855&   \nodata   & $17.73\pm0.03$& $17.19\pm0.02$& $16.38\pm0.03$& 2000 Nov 17 &  \\
               & $18.20\pm0.03$ & \nodata  &  \nodata  &   \nodata  & 2000 Nov 18 & \\ 
               &   \nodata & \nodata  &  \nodata  & $16.81\pm0.03$ & 2001 Oct 5 & $\Delta K = +0.43$ \\ 
Q0642+449  &    \nodata      &$17.18\pm0.04$ &$16.56\pm0.05$ &$15.70\pm0.03$& 2000 Nov 17&  \\
           &  $17.73\pm0.06$ &  \nodata    &   \nodata   &  \nodata   & 2000 Nov 18& \\ 
           & \nodata  & \nodata  & \nodata & $15.75\pm0.06$ & 2001 Oct 5 & $\Delta K = +0.05$ \\ 
%
B30749+426    & $17.16\pm0.05$  &  \nodata &   \nodata &   \nodata  & 2000 Nov 18 &  \\
           &\nodata & $16.48\pm0.02$$^{d}$ & $16.04\pm0.01$$^{d}$ & $15.53\pm0.01$$^{d}$& 2001 Apr 15 & IRCAM \\ 
            &\nodata & \nodata& \nodata & $15.47\pm0.03$& 2002 Feb 16 & $\Delta K = -0.06$\\ 
%
PMNJ0833+0959  &  $18.24\pm0.03$ &   \nodata     &  \nodata     &  \nodata    & 2000 Nov 18 &   \\
               &  \nodata  & $17.69\pm 0.02$$^{d}$ & $17.08\pm 0.02$$^{d}$ & $16.62\pm 0.02$$^{d}$ &2001 Apr 15 & IRCAM \\ 
               &  \nodata &   \nodata     &  \nodata     & $16.61\pm0.02$ & 2002 Feb 16 & $\Delta K \sim0$ \\
\end{tabular}
\end{minipage}
%
%
\begin{minipage}{140mm}
{\it a~}{All magnitudes except those obtained with IRCAM and those obtained 2002 July 17 
were measured within an aperture of radius 10 pixels (1\farcs8 diameter) and an aperture 
correction applied to determine the magnitude within the 45 pixel radius (8\farcs19 diameter) 
aperture used for the standard stars.
The quasar magnitudes obtained using UFTI on 2002 July 17 were measured within apertures of
radius, R=20(3\farcs6 diameter), and corrected to the R=30(5\farcs4 diameter) aperture 
used in determining the zeropoint.  
Magnitudes are determined assuming only an extinction correction; a synthetic color 
correction to account for the different spectral shapes of Vega and the quasar is 
incorporated in the zero-magnitude flux.} \\
{\it b~}{2002 Jul 17 was not strictly photometric and there were not enough stars for 
relative photometry, but the data appeared to be of sufficient
quality to show that neither of these quasars varied appreciably.} \\
{\it c~}{The quasar magnitudes obtained for PC0027+0525 on 2003 Aug 08 (in sub-half-arcsecond seeing) 
were measured within apertures of radius, R=10(1\farcs8 diameter), and corrected to the R=30(5\farcs4 diameter) 
aperture used in determining the zeropoint. Bad pixels were interpolated over before magnitudes were computed; this
affected the J-band magnitude. The quasar was barely detected on the 2000 Nov 17/18 data,
though it was clearly seen on the 2003 Aug 08 images. The Nov 17/18 magnitudes are only listed to
show their rough consistency with the 2003 magnitudes.}\\
{\it d~}{JHK magnitudes for these two objects were measured with IRCAM.
For these observations, aperture corrections were made from an R=20(3\farcs25 diameter) to
R=40(6\farcs5 diameter) aperture.} \\
\end{minipage}
\end{table*}

All of the near-IR observations were carried out in a grid (or jitter) pattern. At J, H
and K, this was either a 5 or 9-point jitter, depending on source 
brightness, of individual 60-second observations. For the faintest sources,
the 9-point jitter was repeated 2 or 3 times. At Z, longer integrations were
needed to be background limited, and most of the Z band observations were 
carried out as 5 or 3-point jitters of 120 or 250-second integrations. 

Standard stars from the set of UKIRT faint standards (Hawarden et al. 2001)
were observed approximately every hour for flux
calibration. While nights in September and October were photometric,
the two in November had some cloud, during the second part of Nov 17 
and first part of Nov 18. Frequent observations of standards
enabled the data from the beginning of Nov 17 to be salvaged, and the 
two quasars observed through cirrus at the end of that night
were re-observed with IRCAM in April 2001. Since there 
was cirrus at the start of Nov 18,
it was treated as non-photometric and repeat observations to build up signal-to-noise 
on several faint objects were made.
Later on, skies cleared and standard star 
observations looked reliable, so Z band observations were made of several 
of the targets.

Reduction of the near-IR data was straightforward up to the point of
doing aperture photometry. The {\sc ORAC-DR} pipeline \footnote{ORAC-DR is an online
Data Reduction Pipeline developed at the Joint Astronomy Centre by Frossie Economou
and Tim Jenness in collaboration with the UK Astronomy Technology Centre as part of the ORAC
project.} 
(Currie et al. 1999)
was used to process the images: basically it (1) subtracts a dark, (2) applies a
bad-pixel mask, (3) combines the 3, 5 or 9 frames from each group to make a flat field, 
masking the central region where the object should fall, (4) divides this 
flat into each observation of the group and finally (5) shifts and 
averages these to produce the final mosaic.

Aperture photometry was done using the {\sc IRAF}
\footnote{IRAF is distributed by the National Optical Astronomy Observatories,
which are operated by the Association of Universities for Research
in Astronomy, Inc., under cooperative agreement with the National
Science Foundation.} 
task {\sc digiphot.apphot.phot}.
The difficulties in aperture photometry arose primarily because of 
greater-than-average and variable seeing.
For the standard stars, a 45-pixel radius aperture was
adopted (diameter $8\farcs2$).  The curves of growth for many of the
quasars became noisy at radii larger than 10 or 15 pixels, however, so
an aperture correction was necessary. For consistency, all of the quasar
data were corrected from a small aperture of radius 10 pixels (diameter 
$1\farcs8$) to the 45 pixels standard star one. The variable seeing meant 
that the aperture correction had to be determined individually for each 
mosaic. 
Many fields had only one star bright enough to use in determining
an aperture correction. In this case, the difference between its instrumental
magnitudes measured within the R=10 pixel and R=45 pixel apertures was
taken to be the aperture correction, $AC$, and the quadrature 
sum of the errors in these instrumental magnitudes was adopted as the error in it.
A few fields had several
stars bright enough for aperture corrections. For these, the
aperture correction was determined from the star with the smoothest curve of
growth and the error was taken to be the standard deviation,
$\sigma_{N-1}$, in the measurements. 
The aperture corrections determined in this way, for each frame, were
reasonable | not too scattered and generally following the seeing 
variations. 
On a single night with good seeing ($\sim0\farcs5$, 2000 October 13), 
average aperture corrections ranged from $-0.18\pm0.06$ at J to 
$-0.12\pm0.01$ at K, while on a night with poor seeing 
($0\farcs8$ to $2\arcsec$, 2000 November 17) they ranged from $-0.52\pm0.17$ at J 
to $-0.41\pm0.11$ at K.
The error here reflects the seeing variations and is smaller for an 
individual frame.

Zero-points, $ZP$, and extinction coefficients, $k_{ext}$, were computed 
for each night at each filter. The quasar magnitudes, $m_{Q}$, were then 
determined from their instrumental magnitudes within the R=10 aperture, $m_{10}$, 
as: 
\begin{equation}
m_{Q} = m_{10} - AC - k_{ext}(\chi - 1) + ZP 
\end{equation}
where $\chi$ is the airmass.
Errors are quadrature sums of the
errors in the instrumental magnitude, aperture correction and zero point.

\subsection{Variability} \label{var}
Prior to these observations, most of the target quasars had not been
extensively observed; in particular, there is little information on their 
variability properties. 
Section \ref{rspec} confirms that most of the radio-selected {\em z3f} quasars have
flat radio spectra. The set of flat-spectrum radio-loud quasars includes blazars, which contain a
beamed synchrotron component that may contribute significantly
to the rest-frame optical (e.g. Serjeant \& Rawlings 1997), thus complicating
the use of the rest-frame optical/UV continuum shape in probing the central mass
and accretion rate.
Signatures of beamed emission include rapid, high-amplitude variability
(e.g. March\~{a} et al. 1996). 
Through service observations at UKIRT,
all 20 targets were re-observed at K, approximately one year after 
the initial run. Conditions were not always photometric, and differential photometry
was done for most of the fields by comparison with the earlier images. 
The resulting magnitudes are listed in Table \ref{z3fir}. 
Three quasars varied by at least 0.4 mag: TXS2358+189, MG2 J062425+3855 and 
PKS0201+113, which brightened by 1.2 mag at K in 11 months. It is likely that 
beamed synchrotron emission contributes to the rest-frame optical continua
of these three, however, the data do not rule out the possibility of such a component
in the other quasars, even though they varied less.
It should be possible 
to select radio-quiet samples from optical multi-color or grism surveys or deep 
near-IR surveys that are not biased to blue 
optical/UV continua and avoid the complications of radio-optical synchrotron contributions.

\section{Spectral Energy Distributions}\label{sseds}

\subsection{Magnitude-to-flux conversion} \label{zeromagflux}
No color-correction was made to the IR magnitudes listed in Table \ref{z3fir}. 
However, a set of zero-magnitude fluxes which incorporate a synthetic color 
correction were computed. These were determined to be the fluxes, at the central
wavelengths of each of the IR bands, of a power law: F$_{\nu,PL} = n~ \nu^{-0.46}$ 
(representative of the shape of the composite spectrum of 657 quasars from the FBQS, 
Brotherton et al. 2001); which has been normalized to have the same broad band magnitude 
as Vega, i.e.  
\begin{equation}
n = {\int{{\phi_{\lambda} F_{\lambda,Vega} d\lambda}} \over \int{{\phi_{\lambda} F_{\lambda,PL} d\lambda}}},
\end{equation}
where $\phi_{\lambda}$ represents the transmission through the filter, multiplied by the 
atmospheric transmission and detector response when available.

Color-corrected zero-magnitude fluxes were computed for each of the broadband
filters used in UFTI. For these, $\phi_{\lambda}$ incorporates both the filter and 
atmospheric transmission functions.
Data from two other instruments (Paper I): the MMT-IR photometer (Rieke 1984) 
and OSIRIS (DePoy et al. 1993) at CTIO; are used to re-compute rest-frame continuum shapes
for the {\em z3b} quasars (section \ref{othersam}). 
For these, $\phi_{\lambda}$ was simply assumed to have a tophat shape starting and ending 
at the 50\% cut-on and cut-off wavelengths of the {\it MKO-NIR} filters.
For each filter and transmission function (UFTI or `tophat'), the central wavelengths and 
color-corrected zero-magnitude fluxes are listed in the second and fourth columns of 
Table \ref{calib}.
For comparison, the zero-magnitude fluxes for Vega (i.e. the flux of Vega at each of the central 
wavelengths), are also included in the Table. 
The zero-magnitude fluxes were found to depend little on the shape of $\phi_{\lambda}$, 
so for simplicity the color-corrected zero-magnitude fluxes computed for UFTI were used 
to convert all the IR magnitudes to fluxes.

\begin{table*}
\centering
\begin{minipage}{140mm}
\caption{Zero-magnitude fluxes}\label{calib}
\begin{tabular}{llll} 
Filter$^{a}$ & $\lambda_{0}^{b}$ & F$_{\lambda}$(Vega)$^{c}$ & F$_{\lambda}$(PL)$^{c}$ \\
UFTI Z         & 0.9525 & 5.239 & 6.921  \\
UFTI MKO-NIR J & 1.25 & 2.982 & 3.040  \\
`tophat' J ($1.17-1.33\mu$m) & 1.25  & 2.982  & 3.021  \\
UFTI MKO-NIR H & 1.635 & 1.192 & 1.208  \\
`tophat' H ($1.49-1.78\mu$m) & 1.635 & 1.192  & 1.203  \\
UFTI MKO-NIR K & 2.20 & 0.4151 & 0.4192  \\
`tophat' K ($2.03-2.37\mu$m) & 2.20  & 0.4151  & 0.4189  \\ 
%
%
%
\end{tabular}
\end{minipage}
\begin{minipage}{140mm}
{\it a~}{Name of the filter. I did not have filter curves for the MMT 
IR photometer or OSIRIS so 
approximate the filter by a tophat and give the adopted cut-on and cut-off values in
the table.} \\
{\it b~}{The central wavelength (\AA) of the filters used in UFTI. Since
these filter curves are close to a tophat shape and the cut-on and cut-off values are set
mainly by the atmosphere, these central wavelengths are used for the MMT and OSIRIS data 
as well.}\\
{\it c~}{Zero-magnitude flux [$10^{-10}$ erg s$^{-1}$ cm$^{-2}$ \AA$^{-1}$]: 
the flux, $F_{\lambda}(\lambda_{0})$, of Vega or of a $\nu^{-0.46}$ power 
law normalized to have the same broad band magnitude as Vega.}\\
\end{minipage}
\end{table*}

\subsection{Galactic extinction} \label{galext}

The fluxes were corrected for Galactic extinction, assuming the
values of E(B-V) from Schlegel, Finkbeiner \& Davis (1998; 
as reported in NED), and the near-IR and optical Galactic reddening law of 
Rieke, Lebofsky \& Low (1985) and Savage \& Mathis (1979).

\subsection{Spectral energy distributions} \label{blueshift}

\begin{figure*}
\vspace*{8.5in}
\includegraphics{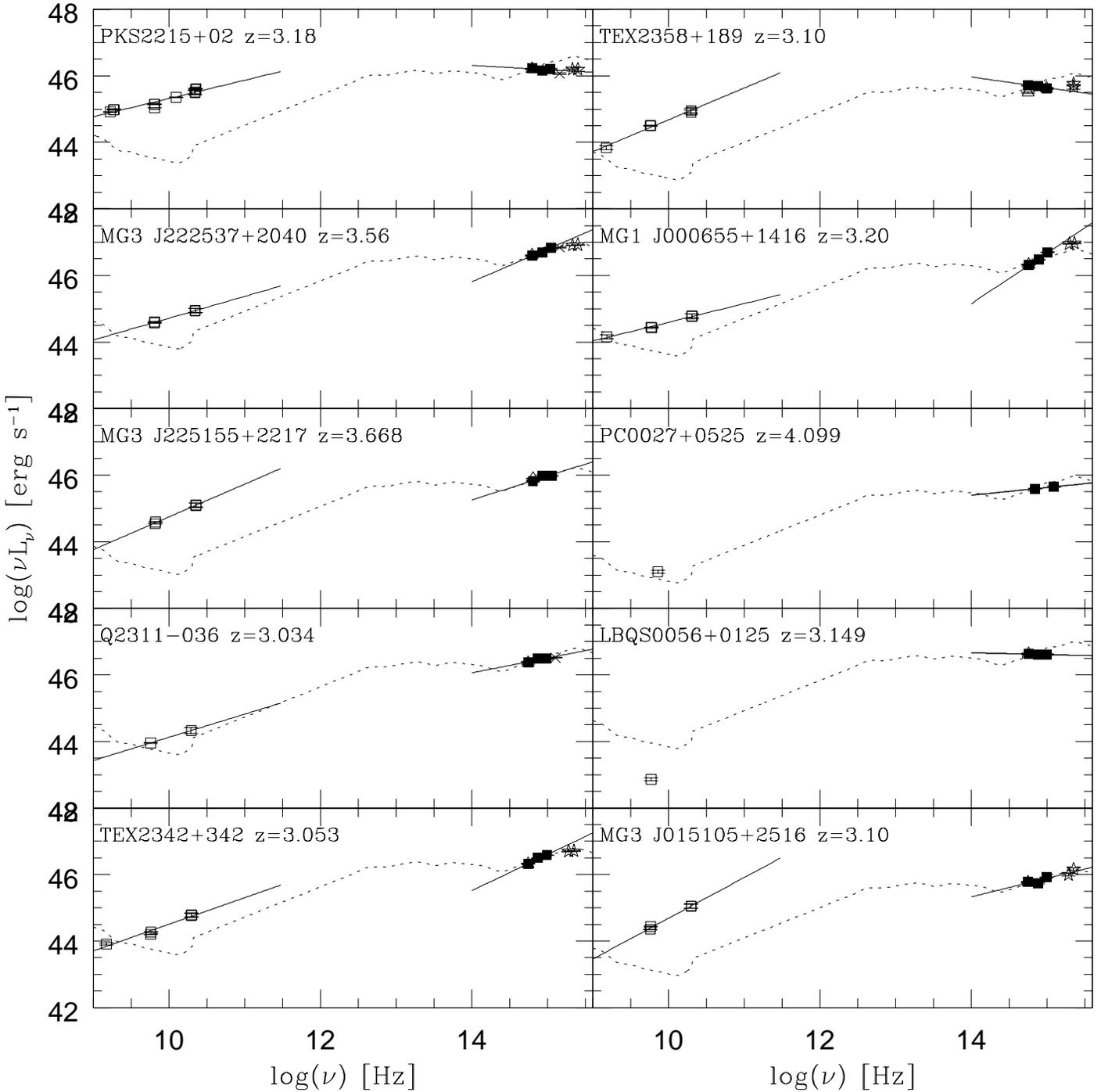} 
  \caption[]{Rest-frame radio/optical/UV spectral energy distributions (SEDs)
  of the 20 `faint $z>3$' sample quasars. Radio data (open boxes) are taken from NED, 
  and the optical (stars; see Appendix A) and IR data (filled boxes for the `original' JHK data; 
open triangles for
  K band `variability check' measurements and a few J datapoints not simultaneous with those used
  in the power law fits; and X's for the Z band points) 
  are from this paper.  The mean SED of $\sim18$ low redshift
  ($z\sim0.1$) radio-loud quasars (dotted line; Elvis et al. 1994a), normalized to the quasar's
  rest-frame 4400\AA\ luminosity, is overplotted
  for comparison. Power law fits to the radio continua and to the rest-frame optical/UV
  continua (from JHK only) are plotted (solid line segments | these extend beyond the
  frequency ranges of the fits so that they would be easy to see).
  \label{seds}}
\end{figure*}

For the 20 {\em z3f} quasars, rest-frame radio-to-UV spectral energy distributions (SEDs) 
were constructed from the data presented here and from radio data from the literature 
(listed in NED). These are plotted in Figure \ref{seds}.

\begin{figure*}
\vspace*{8.5in}
\includegraphics{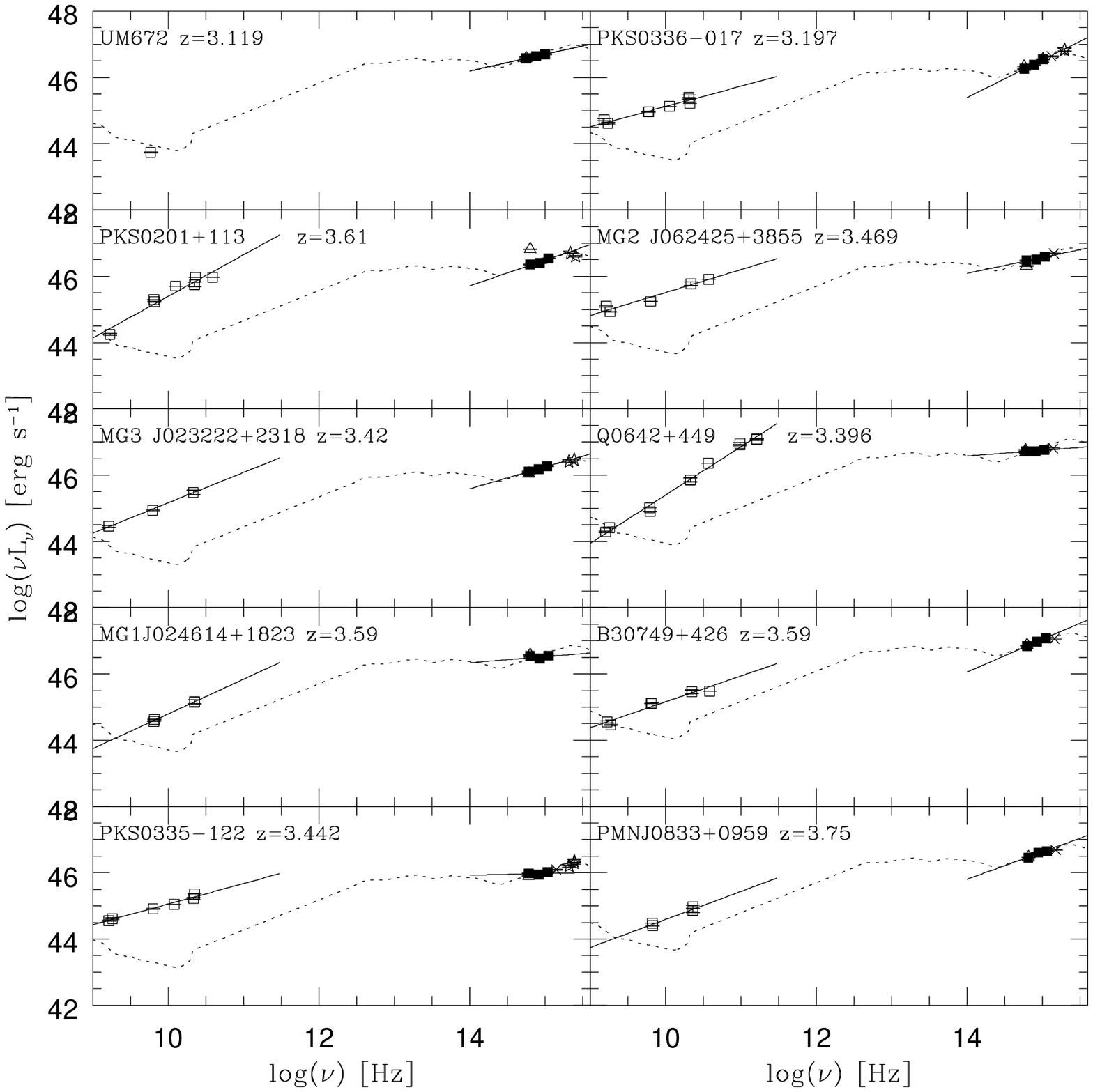} 
\centerline{Figure \ref{seds} | {\it Continued}}
  \end{figure*}

\section{Spectral indices}\label{salphas}
\subsection{$\alpha$ for the faint $z>3$ sample}\label{salphasopt}
\begin{figure}
\vspace*{3.5in}
\includegraphics{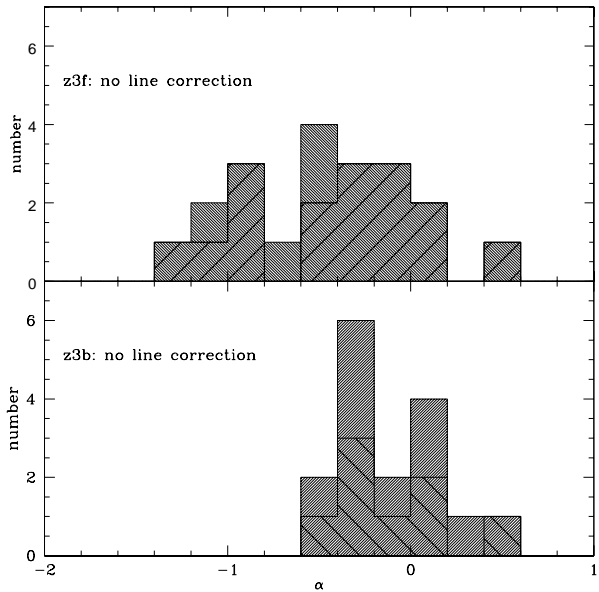} 
\caption{Distributions of rest-frame optical spectral indices for the 
20 {\em z3f} quasars (top) and 16 {\em z3b} quasars (bottom).
Superposed on these are the distributions for the 16 radio-selected {\em z3f} quasars
(top; right slanted thick hash marks) and the 8 radio-loud {\em z3b} ones (bottom;
left slanted thick hash marks).
Note the red tail to the {\em z3f} sample, with 3 of the quasars, 
2 radio-selected and 1 optically-selected, having $\alpha < -1$. 
\label{hist_aopt}}
\end{figure}

Rest-frame optical/UV spectral indices were determined by fitting a power law through the
rest-frame SEDs between 1500 and 5990 \AA\ ($\log{\nu}$=14.7-15.3). This wavelength range 
was chosen since it includes J, H and K for $z=3-4.1$, the range of redshifts in the 
sample. The power law
fits were made with and without the Z-band point, which was not obtained for 
all the objects, but the difference is minimal. For the analysis, the fits to 
just J, H and K were used.
The 1500-5990\AA\ spectral indices determined from these fits (with and without
the emission line correction made to the photometry; see Appendix B) are listed in
Table \ref{alphas} and their distribution
is plotted in Figure \ref{hist_aopt}. 

\begin{table*}
\centering
\begin{minipage}{155mm}
\caption{Rest-frame optical and radio spectral indices}
\label{alphas}
\begin{tabular}{l rc rc rc} 
Quasar & $\alpha_{opt}^{a,b}$ & L$_{4400\AA}^{a,b}$ & $\alpha_{opt}^{a,c}$(corr) & L$_{4400\AA}^{a,c}$(corr)& $\alpha_{r}^{d}$ & L$_{5GHz}^{d}$\\
PKS 2215+02     & $-1.30\pm0.05$ & 46.22 & $-1.24\pm0.05$ & 46.13 & $-0.45\pm0.06$& 45.15 \\
MG3 J222537+2040& $-0.21\pm0.04$ & 46.62 & $-0.16\pm0.04$ & 46.55 & $-0.35\pm0.02$& 44.52 \\
MG3 J225155+2217& $-0.31\pm0.13$ & 45.85 & $-0.34\pm0.13$ & 45.77 & $-0.02\pm0.05$& 44.45 \\
Q2311-036       & $-0.65\pm0.05$ & 46.43 & $-0.75\pm0.05$ & 46.40 & $-0.30\pm0.00$& 43.92 \\
TXS 2342+342    & $-0.08\pm0.04$ & 46.42 & $-0.10\pm0.04$ & 46.38 & $-0.20\pm0.09$& 44.27 \\
TXS 2358+189    & $-1.37\pm0.09$ & 45.70 & $-1.51\pm0.09$ & 45.66 & $-0.05\pm0.07$& 44.39 \\
MG1 J000655+1416& $+0.52\pm0.08$ & 46.41 & $+0.36\pm0.08$ & 46.36 & $-0.44\pm0.03$& 44.43 \\
PC 0027+0525    & $-0.77\pm0.09$ & 45.70 & $-0.94\pm0.09$ & 45.58 & $-0.18^{e}$   & 43.22$^{e}$ \\
LBQS 0056+0125  & $-1.05\pm0.10$ & 46.62 & $-1.19\pm0.10$ & 46.57 & $-0.18^{e}$   & 42.92$^{e}$ \\
UM 672          & $-0.50\pm0.09$ & 46.62 & $-0.66\pm0.09$ & 46.58 & $-0.18^{e}$   & 43.79$^{e}$ \\
MG3 J015105+2516& $-0.44\pm0.09$ & 45.79 & $-0.61\pm0.09$ & 45.75 & $+0.22\pm0.08$& 44.32 \\
PKS 0201+113    & $-0.23\pm0.07$ & 46.36 & $-0.33\pm0.07$ & 46.29 & $+0.25\pm0.13$& 45.02 \\
MG3 J023222+2318& $-0.35\pm0.07$ & 46.13 & $-0.43\pm0.07$ & 46.06 & $-0.08\pm0.03$& 44.89 \\
MG1 J024614+1823& $-0.82\pm0.09$ & 46.49 & $-0.98\pm0.09$ & 46.43 & $+0.05\pm0.06$& 44.48 \\
PKS 0335-122    & $-0.67\pm0.05$ & 45.96 & $-0.57\pm0.05$ & 45.90 & $-0.38\pm0.04$& 44.87 \\
PKS 0336-017    & $+0.08\pm0.05$ & 46.34 & $+0.06\pm0.05$ & 46.28 & $-0.39\pm0.05$& 44.94 \\
MG2 J062425+3855& $-0.40\pm0.04$ & 46.48 & $-0.34\pm0.04$ & 46.41 & $-0.31\pm0.08$& 45.30 \\
Q0642+449       & $-0.79\pm0.06$ & 46.72 & $-0.77\pm0.06$ & 46.65 & $+0.46\pm0.05$& 44.96 \\
B3 0749+426     & $-0.11\pm0.03$ & 46.88 & $-0.13\pm0.03$ & 46.81 & $-0.22\pm0.08$& 44.93 \\
PMN J0833+0959  & $-0.38\pm0.04$ & 46.50 & $-0.33\pm0.04$ & 46.44 & $-0.16\pm0.11$& 44.34 \\
\end{tabular}
\end{minipage}
\begin{minipage}{155mm}
{\it a~}{Power law fit through $\log\nu_{rest}=14.7-15.3$ ($\lambda_{rest}\sim1500-5990$\AA) 
and monochromatic rest-frame luminosity [erg s$^{-1}$] of this power law fit at 
$\lambda_{rest}=4400$\AA.}\\
{\it b~}{No correction has been made for emission lines}. \\
{\it c~}{Emission line contribution has been estimated and removed (see Appendix B)}. \\
{\it d~}{Power law fit through $\log\nu_{rest}=9-11.5$ and luminosity [erg s$^{-1}$] of 
this power law fit at $\nu_{rest}=5$ GHz.} \\
{\it e~}{Only one radio data point was found, from NVSS and at 1415MHz. To estimate the
rest-frame luminosity at 5GHz, the median spectral index determined for the 16 radio-selected
quasars, $\alpha_{r} = -0.18$, was assumed.} \\
\end{minipage}
\end{table*}

\subsection{Bright $z>3$ sample}\label{othersam}

\subsubsection{$\alpha$ for the bright $z>3$ sample}

The original aim of this project was to expand, in number and luminosity range,
the set of 15 bright $z>3$ quasars for which Kuhn et al. (2001) constructed rest-frame 
optical/UV SEDs (Paper I, table 1). In this section, the SEDs of the new {\em z3f} quasars are 
compared with those of the bright {\em z3b} ones, to determine whether increasing the number 
has reduced the scatter, and whether there is any trend with luminosity. 
The first step in making the comparison is to re-determine the spectral indices of the
{\em z3b} quasars following the same methods as described in sections \ref{sseds} and 
\ref{salphasopt} for the {\em z3f} sample,
namely from the broad band magnitudes alone. Dropping the requirement for IR spectroscopy
adds 2 quasars to the {\em z3b} sample, SP82 1 and Q1442+101, for which we had obtained near-IR magnitudes 
(table 6 of Paper I) but failed to obtain spectra. One quasar which was missing 
photometry at J (Q0114-089) had to be dropped. This yields a net total of 
16 quasars in the revised {\em z3b} sample.
Finally, I observed the lensed quasar, Q1208+101, on 21 May 2001 at UKIRT and 
obtained the following magnitudes (for both components): Z=$16.60\pm0.01$, J=$16.09\pm0.01$, 
H=$15.63\pm0.01$ and K=$15.16\pm0.01$.
At J and K, these are 0.15 and 0.26 magnitudes brighter than what Kuhn et al. (2001) 
measured in 1993; the differences are within range of what is observed for quasar
variability (Giveon et al. 1999; Hook et al. 1994 remark on the anomalously high variability
of Q$0055-269$, another quasar from the {\em z3b} sample that was discovered in a similar 
manner as Q$1208+101$). In this paper, only these new data are used for Q1208+101.

For the quasars that had been observed with OSIRIS, one photometric point was sufficient 
to normalize the entire $1-2.5\mu$m cross-dispersed spectrum, so we had not obtained 
photometry at all three 
bands: J, H and K.  To derive broad band magnitudes from the spectra, I integrated over the 
product of these with the curves of filter-plus-atmospheric transmission (tophat with cut-on 
and cut-off values as listed in Table \ref{calib}), first, to determine the scaling factor 
needed to match the integral with the broad band photometry (typically at H), and second, 
to determine magnitudes in the bands for which we had not obtained photometry. 

The near-IR magnitudes, determined from the OSIRIS spectra or directly, for all 
16 {\em z3b} quasars were processed in the same way as described above for the {\em z3f} quasars: 
they were converted to fluxes (section \ref{zeromagflux}), 
dereddened by the Galactic value (section \ref{galext}),
corrected for emission line contributions (Appendix B) and blueshifted 
to the rest-frame (section \ref{blueshift}). 

\subsubsection{Broad-band photometry vs. Narrow band luminosities}

\begin{figure}
\includegraphics{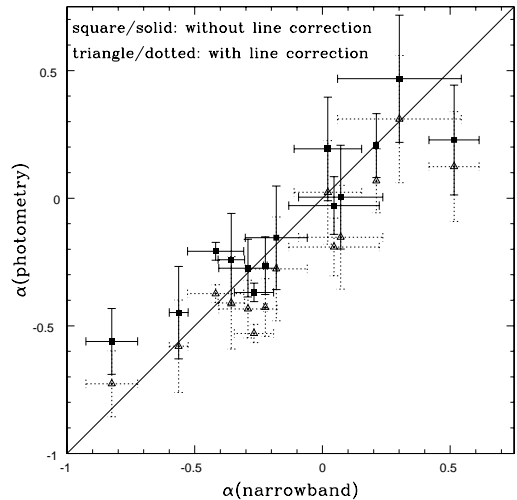} 
\vspace*{6.5cm}
  \caption{Comparison of the optical spectral indices measured from fits to broad 
  band photometry, with (open triangles and dotted error bars) and without (filled squares and solid 
  error bars) a correction for emission line contributions, 
  against those determined from narrow band monochromatic luminosities (Paper I, Table 15). 
  The line, $\alpha$(photometry) = $\alpha$(narrowband), is drawn for reference.
  \label{aopt_compare} }
\end{figure}

To confirm the validity of using broad band magnitudes
to measure rest-frame optical spectral indices, for 14 of the 15 bright $z>3$ quasars studied
in Paper I (all but Q$0114-089$), the optical spectral
indices computed from the photometry alone (this time using the zero-magnitude fluxes listed
in Table 9 of Paper I) were compared with those determined by fitting a line through
a set of narrow band monochromatic luminosities (Paper I, Table 15) which covered approximately 
the same rest-frame region as did J, H and K: i.e. 3023\AA, 4200\AA, 4750\AA\ and 5100\AA\ for 
$z=3-3.5$; 
2660\AA, 3023\AA, 4200\AA\ and 4750\AA\ for $z=3.6-3.8$; and
2500\AA, 2660\AA, 3023\AA\ and 4200\AA\ for $z>4$.
Figure \ref{aopt_compare} compares the optical spectral indices determined by these two methods
and shows that, while not in perfect agreement, the distributions of spectral indices computed
by fitting through narrow-band line-free windows and through only broad band photometry
do not differ significantly. As can be seen in Figure \ref{aopt_compare}, making the emission line 
correction (Appendix B) improves the agreement for some objects but worsens it for others.

\subsection{Comparing the bright and faint}

As the histograms in Figure \ref{hist_aopt} show, the {\em z3f} sample includes redder
quasars than the {\em z3b} one. Three of the {\em z3f} quasars have spectral indices
$\alpha<-1$ and are ``red quasars'' according to the definition of Gregg et al. (2002). 
A Kolmogorov-Smirnov (K-S) test gives a probability of
5\% that the distributions for the {\em z3b} and {\em z3f} samples are drawn from the same 
parent population. A Student's t-test
shows that the mean spectral indices for each sample differ significantly | the probability
of randomly obtaining a separation as large as measured, $P(t)$, is $1.4$\%.

A number of explanations can be put forth: 

First, the {\em z3f} sample contains
only radio-loud quasars whereas the {\em z3b} sample is evenly divided between radio-quiet
and radio-loud. Several of the {\em z3f} quasars varied, which suggests that they
may be blazars in which a non-thermal contribution to the rest-frame optical could 
redden their rest-frame optical/UV continua (Whiting et al. 2001). If the two radio-selected 
quasars in the red tail are removed from the {\em z3f} sample, the K-S probability that the 
bright and faint samples are drawn from the same parent population increases to 13\% | there are no
longer sufficient objects in the tail to render the distributions of spectral indices
significantly different. 
However the mean spectral indices for each sample are still significantly different ($P(t)=4$\%).

\begin{figure*}
\vspace*{5in}
\includegraphics{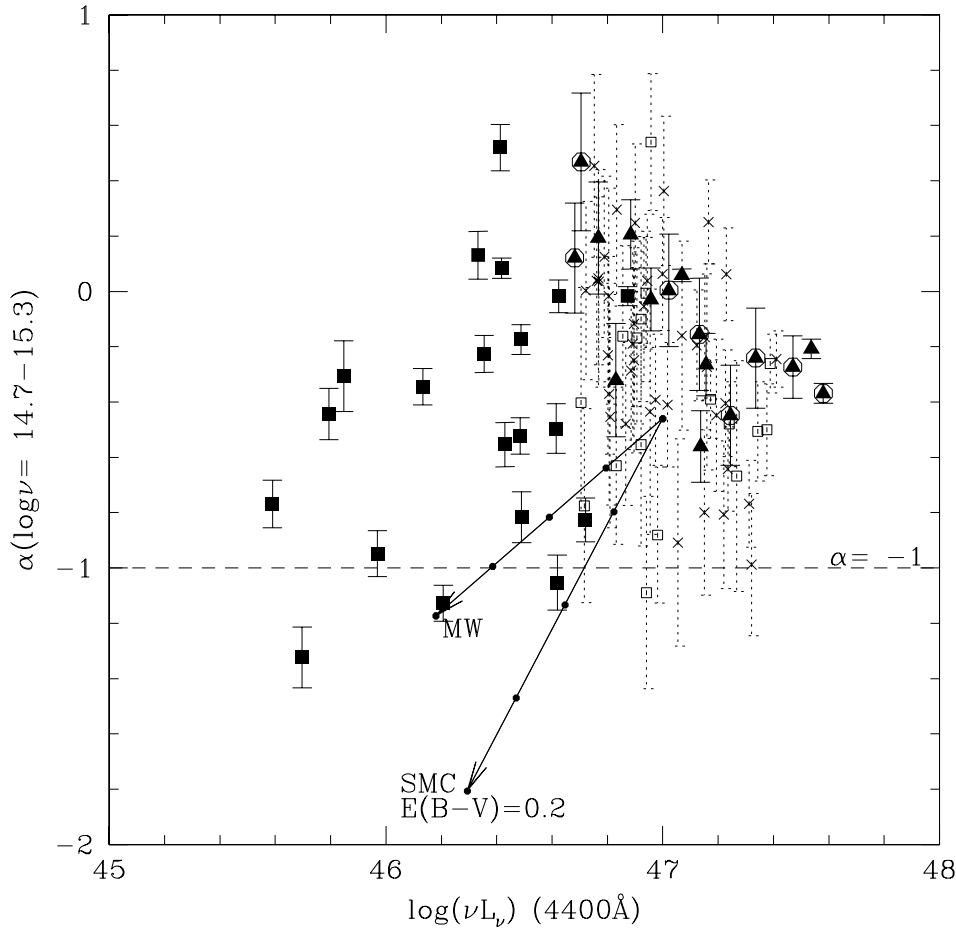}
  \caption{Rest-frame optical spectral index, $\alpha$, {\it vs.} luminosity 
  for several samples: the 20 {\it z3f} 
  sample quasars (filled boxes); the 16 quasars in the {\it z3b} 
  sample (filled triangles; 
  the radio-loud ones are circled); and 56 $z>3$ quasars, 17 radio-loud (open 
  boxes) and 39 radio-quiet (crosses) from VCV9 
  for which J, H and Ks magnitudes were published in the 2nd incremental data release 
  of the 2MASS PSC. 
  Starting at $\alpha=-0.46$, reddening vectors are plotted for E(B-V) = 0 to 0.2 in 
  steps of 0.05 and assuming both Galactic and Small Magellanic Cloud 
  extinction laws.
  \label{slum}}
  \end{figure*}

Second, dust reddens and extincts the emitted continuum, so in the optical/UV, 
redder objects would be expected to be fainter. Richards et al. (2001) 
note this trend in the sample of $\sim2600$ SDSS quasars for which they present 
optical photometry, and it is evident in Figure \ref{slum} which plots spectral index
against optical luminosity for the faint and bright $z>3$ samples as well as the 56
$z>3$ quasars with J, H and Ks magnitudes from the 2MASS 2nd incremental data release.

Finally, the redder colors of the {\em z3f} with respect to the {\em z3b} quasars
could represent an intrinsic difference in their central engines. 
The spectrum emitted by a thin accretion disk is expected to peak at a frequency determined
by the apparent disk temperature, which for optically thick/geometrically thin disks follows the
relation: $\log T \sim {1\over4} \log{\dot{m} \over M} - 2.4(\cos\theta-1)$ where $\dot{m}$ is the 
accretion rate relative to the Eddington limit, $M$ is the black hole mass, and $\theta$ 
is the inclination of the
disk axis to the line of sight (Sun \& Malkan 1989, McDowell et al. 1991).
The thermal disk spectra emitted by the fainter set would be systematically redder if, for 
example, both sets of quasars had similar central masses, but the fainter ones were accreting 
at lower rates.

The importance of dust at high redshift (e.g. Andreani, Franceschini \& Granato 1999; 
Warren, Hewett \& Foltz 2000), together with the possibility of a 
significant non-thermal component in these radio-loud quasars, leads to the suspicion 
that extrinsic factors are responsible for the on-average redder colors of these {\em z3f}
quasars. 
These complicate the use of optical/UV continua to probe the central engines and point
to the need for a similar study of radio-quiet quasars and for spectroscopic data. 

\section{Red quasars}\label{red}

Three of the sample quasars have $\alpha < -1$; two of these, 
PKS2215+02 and TXS2358+189, are radio-selected, and one, LBQS 0056+0125, was discovered 
in a grism survey (PC0056+0125, Schmidt, Schneider \& Gunn 1987).

\subsection{LBQS 0056+0125}
LBQS0056+0125 has an optical spectral index $\alpha=-1.05$ (Table \ref{alphas}).
In the discovery paper, Schmidt, Schneider \& Gunn (1987) remarked on the presence of 
narrow absorption lines redwards of Ly$\alpha$.
Since then several line-of-sight Ly$\alpha$ absorption line systems have been detected; one
at $z=2.7771$ (Schneider, Schmidt \& Gunn 1991; Pettini et al. 1997) is damped (DLA).
To redden an average quasar continuum (the FBQS composite)
emitted at the quasar redshift, $z=3.149$, to $\alpha=-1.05$, requires an extinction 
E(B-V) $\sim 0.12$ in the DLA, assuming the Small Magellanic Cloud (SMC)
extinction law (Pr\'{e}vot et al. 1984, Bouchet et al. 1985).
Pettini et al. (1997) measure the neutral hydrogen column for this DLA to be 
$\log(N_{HI})=21.11\pm0.07$. The above estimate for E(B-V) is consistent with this if
the gas-to-dust ratio in the DLA is approximately 2 times that in the Milky Way(MW; $4.8\times10^{21}$
cm$^{-2}$ s$^{-1}$, Savage \& Mathis 1979) | a 
factor $\sim5-15$ times lower (dustier) than existing data imply for DLAs (Pei, Fall \& Bechtold 1991, 
Pettini et al. 1997).
The above assumes no intrinsic extinction. If the SMC law is used also to describe the extinction in the
host galaxy, and the extinction in the DLA is fixed at E(B-V)=0.027, what is implied by $\log(N_{HI})=21.11$
and a gas-to-dust ratio 10 times that in the MW, then an emitted continuum (FBQS composite) 
would need to undergo an intrinsic extinction, E(B-V)$\sim0.1$ mag, which corresponds
to an intrinsic column density N$_{HI}\sim 5\times10^{20}f$ cm$^{-2}$ where
$f$ is the ratio of the gas-to-dust ratio in the quasar host to that in the MW. 
This is larger than the column densities measured for low redshift optically selected
quasars (Laor et al. 1997), but consistent with the values inferred for sets of X-ray selected 
(Puchnarewicz et al. 1996) and high redshift radio-loud quasars (e.g. Elvis et al. 1994b).
So the red color of LBQS0056+0125 can be explained
by intrinsic and line-of-sight extinction, but with gas-to-dust ratios that 
are on the low side for the intervening absorbers and quasar host galaxies. 
The emitted continuum may itself be redder than the FBQS composite.
Also, the gas-to-dust ratios in DLAs do show a lot of scatter (Pettini et al. 1997); 
this might also explain why other quasars from the {\em z3f} sample with high redshift 
DLAs (e.g. TXS 2342+342 and PKS0201+113) do not also appear very red. 
Warren et al. (2001) have recently imaged the field of LBQS0056+0125 with NICMOS(NIC2) on
HST and detect two possible candidates for the galaxy counterpart to the DLA. 

\subsection{PKS2215+02 and TXS2358+189}
The other two quasars with $\alpha<-1$ are radio-selected. They
have some of the the highest radio-to-optical flux ratios of the 
{\it z3f} sample (see Table \ref{z3fsample}).
PKS2215+02 belongs to the Parkes Half Jansky sample (Drinkwater et al. 1997),
which was found to contain sufficient numbers of red quasars to lead to the suggestion,
later proved well-founded at least for low redshift AGN (Cutri et al. 2000), that 
optical surveys were missing a large fraction of quasars (Webster et al. 1995).
The Parkes Half Jansky sample has been 
well studied in the optical and IR (Francis et al. 2000; 
Whiting et al. 2001) and X-ray (Siebert et al. 1998).
Francis et al. (2000) measured the following magnitudes for PKS2215+02 in early September 
1997: 
B=$21.84\pm0.31$;
V=$20.42\pm0.10$;
R=$20.14\pm0.12$;
I=$20.00\pm0.20$;
J=$19.20\pm0.50$;
H=$18.21\pm0.29$;
K$_{n}$=$19.34\pm1.78$.
These are consistent with the measurements presented here, which were made 3 years later, 
except at J and K\footnote{Though Francis et al. used the K$_{n}$ filter, they normalized
their magnitudes to the K band zero point.}, although they quote a large uncertainty in the
K$_{n}$ magnitude. Between September 2000 and October 2001 the quasar did not vary at K.
In 1995 May/June, PKS2215+02 was detected in a 14.7 ksec observation 
with the ROSAT HRI, with an unabsorbed $0.1-2.4$ keV flux (observed frame)
equal to 3.2$\times10^{-13}$ 
erg s$^{-1}$ cm$^{-2}$  (Siebert \& Brinkmann 1998). 
Siebert \& Brinkmann noted that this quasar is optically very faint and has
the highest X-ray-to-optical and radio-to-optical flux ratios of the radio-loud
$z>3$ quasars they studied. They suggest that it is heavily absorbed although note
that in the optical/UV, PKS 2215+02 does
not stand out as redder than the other Parkes Half Jansky quasars that Francis et
al. (2001) studied. 
The rest-frame optical continuum presented here is
slightly redder than but not inconsistent with that of Francis et al. (2001). 
Ellison et al. (2001) find no evidence for a damped Ly$\alpha$ system 
along the line-of-sight to PKS 2215+02, so if the red color derives from extinction,
this must occur at the quasar. 
The data presented here, which show the quasar as one of the strongest radio emitters of
the {\em z3f} sample and as having one of the highest radio-to-optical flux ratios ($RL=4.6$;
Table \ref{z3fsample}), 
together with evidence for variability at 5GHz (Siebert \& Brinkmann 1998) and in the 
rest-frame optical, favor a non-thermal contribution rather than 
extinction as the cause of the red color. Further monitoring, or other tests for 
blazar-like activity, are needed to confirm this hypothesis.

TXS2358+189 ($z=3.10$), like PKS2215+02, has a very high radio-to-optical flux 
ratio, $RL=4.5$. It has a flat radio 
spectrum, but insufficient data to measure radio variability. At K,
TXS2358+189 dimmed by 0.55 magnitudes from September 2000 to October 2001; 
this was the second-largest one-year variation seen in the sample and 
suggests the presence of beamed synchrotron emission. 
White, Kinney \& Becker (1993) find evidence in its spectrum for a 
Ly$\alpha$ absorption system, though the data do not conclusively
show whether it is damped. 
In sum, it is probable that the red optical continua of both TXS2358+189 and
PKS2215+02 derive from the blending of a synchrotron component.

\subsection{Characteristics of the $\alpha<-1$ quasars}
It is not too surprising to find several red quasars among the {\em z3f} sample.
First, if the $z=2.5 - \sim4$ decline in quasar space density signals the formation epoch of quasars,
then some of the {\em z3f} quasars may be intrinsically young.
If a ULIRG (e.g. Sanders et al. 1989) or a dust-enshrouded phase (e.g. Egami et al. 1996;
Fabian 1999) is the first stage in the evolution of quasars, then
young quasars might be expected to be red (but see Yu \& Tremaine 2002).
Second, the largely radio-selected sample included objects with high radio-to-optical flux ratios
and with flat radio spectra, several of which, from their variability, appear to be blazars
and thus may be reddened by a non-thermal contribution to the rest-frame optical (Whiting et al. 2001).
And third, the probability of DLAs and Ly$\alpha$ systems along the line of sight increases with
redshift. 

The 3 $\alpha<-1$ $z3f$ quasars can be interpreted as reddened by line-of-sight dust or
as containing a strong synchrotron component.
Ongoing and planned optical, near-IR and deep X-ray surveys will
be more sensitive to reddened quasars than previous surveys; here we compare the 
properties of these 3 $\alpha<-1$ quasars with the limits of such surveys as the SDSS,
WFCAM/UKIDSS and Chandra deep fields. 

The 3 $\alpha<-1$ $z3f$ quasars are relatively faint: 
LBQS0056+0125 has r=18.9 (Schneider et al. 1991) and K=15.7;
PKS2215+02 has V=20.6 and K=17.1 and TXS2358+189 has V$\sim21.3$ and K=18.4. 

From the SDSS commissioning data, Fan et al. (2001b) selected a uniform sample of high redshift 
quasars, setting a flux limit at $i^{*}\sim20$; this would have missed one or two of the 
$\alpha<-1$ {\it z3f} quasars.
%
In the near-IR, the LAS (large area survey), which should complement the SDSS, will reach to K=18.4
(http://www.ukidss.org/surveys/surveys.html) so would have picked up all (TXS2358+189 lies at
its limit).
%
The Chandra deep ($\sim 0.1-1$ Ms) fields (Barger et al. 2001; Hornschemeier et al. 2001; Giacconi 
et al. 2002) reach down to flux levels of order $6\times10^{-17}$ to $4\times10^{-15}$ 
erg s$^{-1}$ cm$^{-2}$ ($2-10$ keV). Few of the $z3f$ quasars have X-ray data, but the 
two detected by
ROSAT-HRI have $F_{x}\sim3-6\times10^{-13}$ erg s$^{-1}$ cm$^{-2}$ and the one upper limit
was consistent with a flux, $F_{x}\sim10^{-14}$ erg s$^{-1}$ cm$^{-2}$ 
(Siebert \& Brinkmann 1998). The unabsorbed $0.5-2$ keV (observed frame) 
X-ray fluxes for a set of $z>4$ quasars range from 
$1-30\times10^{-15}$ erg s$^{-1}$ cm$^{-2}$ (Vignali et al. 2001), so 
future deep X-ray surveys should turn up more quasars similar to those in the $z3f$ sample.

\section{Conclusions}

This paper presents near-IR photometry for 20 radio-loud (16 radio-selected) $z>3$ 
quasars.

Some of the targets had not previously been extensively observed | NED lists references only 
to the radio surveys in which they were detected and the source of the optical identification | 
so the photometry presented here are the first for these.

While the underlying aim is to look for evolution in the central engines, 
uncertainties due to extinction and the possible contribution of a non-thermal component 
complicate efforts to estimate key central engine parameters from these data alone.

The main conclusions that can be drawn from this dataset are:

\begin{enumerate}
\item First, the continuum shapes of faint radio-loud $z>3$ quasars show more scatter
than those of the bright $z>3$ quasars from Paper I. Choosing objects with
luminosities close to $L^{*}$, rather than the extremely high
luminosity quasars of the bright $z>3$ sample, has not led to a
convergence in the optical/UV continuum shapes of $z>3$ quasars.

\item And second, the distribution of continuum shapes of these faint radio-loud $z>3$ quasars is
slightly redder than that of the bright $z>3$ ones, which were a mix of radio-loud and 
radio-quiet quasars, with 3 having $\alpha<-1$.  
These 3 are not all radio-selected, so while a synchrotron contribution 
cannot be ruled out as a cause for the red colors of the radio-selected quasars, the 
red optically-selected quasar probably suffers intrinsic or line-of-sight reddening. 
Study of a carefully selected set of high redshift radio-quiet quasars would eliminate the
possibly of synchrotron reddening. Near-IR and optical spectroscopy
would complement the more easily obtained photometry, enabling measures of the 
central mass via the H$\beta$ (e.g. Laor 1998), MgII (McLure \& Jarvis 2002) or CIV (Vestergaard 2002) 
emission line widths, and an estimate of intrinsic reddening via the Ly$\alpha$/H$\beta$ 
ratio (Netzer et al. 1995, Bechtold et al. 1997). 

\end{enumerate}
The 3 $\alpha<-1$ quasars may represent a bridge to a larger population of high redshift 
quasars that are either intrinsically reddened or reddened by line-of-sight systems 
(Warren et al. 2000), the discovery of which is anticipated from
ongoing and future optical, near-IR, and deep X-ray surveys.

\section*{Acknowledgments}
The bulk of the data presented here were obtained at UKIRT, the United Kingdom Infrared
Telescope, which is operated by the Joint Astronomy Centre on behalf on the U.K. Particle
Physics and Astronomy Research Council. Some of the data were obtained as part of the 
UKIRT Service Programme. I thank the staff at UKIRT for their help with
obtaining and reducing the data, in particular Andy Adamson, Malcolm Currie, Chris Davis, Tom Kerr, 
Sandy Leggett and Thor Wold and John Davies for obtaining the optical photometry presented here.
Also I thank the referee, Dr. Stephen Warren, for comments which helped me to improve the paper.
Construction of the optical/UV continua for the bright $z>3$ dataset used a 
large suite of instruments and telescopes, as described in Paper I. One of the instruments 
mentioned in this paper, OSIRIS, is
the Ohio State Infrared Imager/Spectrometer, and was built using funds from NSF award AST 90-161112
and AST 92-18449 by the OSU Astronomical Instrumentation Facility.
This publication makes use of data products from the Two Micron All Sky Survey, which is a 
joint project of the University of Massachusetts and the Infrared Processing and Analysis 
Center/California Institute of Technology, funded by the National Aeronautics and Space Administration 
and the National Science Foundation.
This research has also made use of the NASA/IPAC Extragalactic Database (NED) which is
operated by the Jet Propulsion Laboratory, California Institute of Technology,
under contract with the National Aeronautics and Space Administration.

\appendix
\section[]{Optical photometry}

Optical, V and R-band, photometry for several of the target quasars
was obtained by J. K. Davies over the course of his 5 night run
(2000 August 24-28 UT) at the Jacobus Kapteyn Telescope in La Palma. 
Since these data were not obtained for all of the target quasars,
they were not used in the analysis. However they are plotted in Figure \ref{seds},
which shows that, except for the reddest quasars with $\alpha<-1$, the optical data generally
trace the UV turnover that is seen in the low redshift mean SED (Elvis et al. 1994a).

The optical images were reduced using the {\sc ccdproc} and {\sc apphot} packages
in {\sc IRAF}. All frames were bias subtracted, and
master sky and dome flats were produced for each filter. The bias-subtracted images
were first divided by the sky or dome flats to remove small scale features, and
then, to reduce the residual gradient, they were divided by a heavily
smoothed median of the sky- or dome-flattened science frames.
Standard stars from Landolt (1992) were observed for flux calibration.
Zero points, extinction and color coefficients were determined from
fits to standard star instrumental magnitudes measured through an aperture
of radius 20 pixels (diameter 13\farcs2).
As with the near-IR photometry, aperture corrections were needed.
In this case, the radius of the
small aperture was 6 pixels (diameter 4\arcsec) except for one instance when
it had to be decreased
to 4 (diameter 2\farcs6), as noted in Table \ref{z3fopt}.
The aperture corrections were determined for each image. The telescope jumped
during some of the integrations, causing E-W elongated images with multiple peaks.
These are noted in Table \ref{z3fopt}. The distance between the peaks was comparable to
or smaller than the 20-pixel radius aperture, and the magnitudes were measured
in the same way as from the round images, however the results should be
regarded with more uncertainty than the error in table \ref{z3fopt} indicates.
Since some objects were observed either at V or R only, two sets of transformation
equations, with and without the color terms, were fit. The appropriate set was
used to determine the magnitudes listed in Table \ref{z3fopt}. For the two objects observed
only at V or R, a synthetic color correction was incorporated in the magnitude-to-flux
conversion, by using the zero magnitude fluxes for a $\nu^{-0.46}$ power-law rather than 
for Vega (as discussed in Section 4.1). Effective wavelengths and zero-magnitude fluxes
for Vega and for the $\nu^{-0.46}$ power-law were computed for both the Harris V and R
filters used at the JKT. For V, these are 0.545$\mu$m, $36.46\times10^{-10}$ and 
$36.11\times10^{-10}$  
erg s$^{-1}$ cm$^{-2}$\AA$^{-1}$, respectively. For R, these are
0.639$\mu$m, $22.44\times10^{-10}$ and $22.11\times10^{-10}$ erg s$^{-1}$ cm$^{-2}$\AA$^{-1}$, 
respectively.  
\begin{table*}
\scriptsize
\centering
\begin{minipage}{140mm}
\caption{Optical observations}\label{z3fopt}
\begin{tabular}{lllll} 
Quasar & V$^{a}$ & R$^{a}$ & UT date & comments \\
%
PKS2215+02   & $20.58\pm0.14^{b}$ & $20.19\pm0.05$ & 2000 Aug 28 & tel jumped 1x during V image \\ 
MG3J222537+2040 & $18.78\pm0.03$ & $18.42\pm0.02$ &2000 Aug 25 & \\ 
TXS2342+342& $18.89\pm0.04$ & $18.52\pm0.03$ &2000 Aug 24 & \\ 
TXS2358+189 & $21.45\pm0.19$ & \nodata & 2000 Aug 28 & tel jumped 1x, not color corrected \\ 
            & $21.21\pm0.16$ & \nodata & 2000 Aug 28 & tel jumped 2x, not color corrected \\ 
%
MG1J000655+1416 & $18.49\pm0.04$ & $18.12\pm0.03$ &2000 Aug 26 & tel jumped 1x during R image\\ 
MG3J015105+2516 & $20.46\pm0.09$ & $20.45\pm0.09$ & 2000 Aug 25 & \\ 
PKS0201+113 & $19.90\pm0.05$ & $19.23\pm0.03$ & 2000 Aug 27& \\ 
MG3J023222+2318 & $20.12\pm0.07$ & $19.81\pm0.06$ & 2000 Aug 25& \\ 
PKS0335-122 & $20.15\pm0.10$ & $20.13\pm0.08$ & 2000 Aug 28 &               \\ 
            & $20.30\pm0.08$ &                & 2000 Aug 28 & \\ 
PKS0336-017  &  \nodata & $18.53\pm0.08$ &2000 Aug 28 & not color corrected\\
             &  \nodata & $18.50\pm0.07$ &2000 Aug 28 & not color corrected\\
\end{tabular}
\end{minipage}
\begin{minipage}{140mm}
{\it a~}{All magnitudes were measured within an aperture of radius 6 pixels (4\arcsec
diameter) except PKS2215+02 at R, and an aperture correction was applied to determine the magnitude 
within the 20 pixel radius (13\farcs2 diameter) aperture used for the standard stars. Magnitudes 
are determined assuming an extinction and color correction except when the quasar was observed in
only one band, in which case a synthetic color correction is incorporated into the magnitude 
to flux conversion.}
{\it b~}{On the R-band image, the aperture correction had to be made from an aperture of
radius 4, rather than 6, pixels.}
\end{minipage}
\end{table*}

\section[]{Estimated emission line contributions to the broad band fluxes}

\begin{figure}
\includegraphics{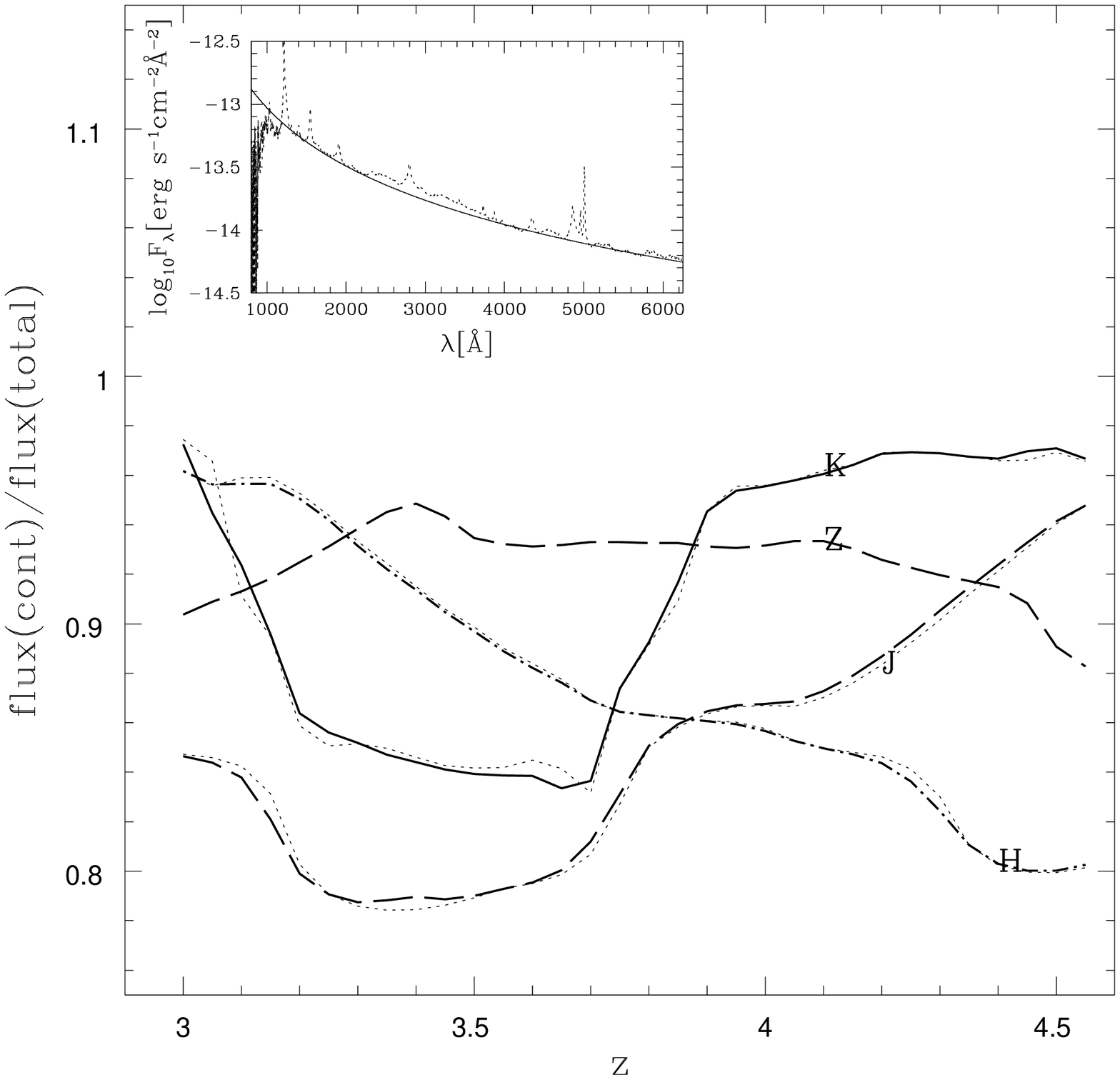}
\vspace{6.5cm}
\caption[]{Ratio of continuum to total (continuum plus emission line) flux
within the Z, J, H and K bands as a function of redshift.
The thick lines are used for the UFTI Z, J, H and K filters
and thin dotted lines for the `tophat' approximations to J, H, K.
Inset is the mean FBQS composite of Brotherton et al. (2001; dotted line)
and continuum fit (a $F_{\nu}\sim\nu^{-0.46}$ power law; solid line)
that were used in this calculation.
\label{linecorr}
}
\end{figure}

For $z=3-4$, the emission lines of MgII, H$\beta$,
[OIII]4959,5007 as well as the blended FeII lines fall within the Z, J, H and K
bands. 
I used the FBQS composite spectrum of Brotherton et al. (2001) to compute, as a function of
redshift, the relative contribution of emission lines
to the continuum flux within the band. The bands were defined as discussed in section 4.1
and listed in Table \ref{calib}. Following Brotherton et al. (2001),
the continuum level was taken to be a power law fit through the composite at
1285, 2200, 4200 and 5770\AA, which had a spectral index, $\alpha= -0.46$.
The ratios of continuum to total flux within a band range from $\sim0.7-0.96$ and 
are plotted as a function of redshift in Figure \ref{linecorr}. For $z>3$, the 
blended FeII and Balmer continuum falls within J and its contribution to the broad
band flux is greater than the emission line contribution at K, which explains
why the spectral indices computed from line-corrected fluxes are redder than those 
computed from uncorrected fluxes (Table \ref{alphas}). 
Differences in line strengths from object to object (e.g. McIntosh et al. 1999, Forster et al. 2001), 
however, are a significant source of uncertainty. 
In the analysis, I have opted to use uncorrected spectral indices since these
involve one fewer assumption and more easily compared to results from other studies.

\bsp

\label{lastpage}

\end{document}